\documentclass[twocolumn,prd,nofootinbib,showpacs,preprintnumbers]{revtex4}
\usepackage{epsfig}

\def\cA{\mathcal{A}}
\def\cP{\mathcal{P}}
\def\cD{\mathcal{D}}
\def\cS{\mathcal{S}}

\begin{document}
\preprint{CERN-PH-TH/2009-007}
\title{Possible causes of a rise  with energy of the cosmic ray positron fraction}

\author{Pasquale D.~Serpico}
\affiliation{Physics Department, Theory Division,
CERN, CH-1211 Geneva 23, Switzerland}

\date{\today}

\begin{abstract}
Based on general considerations rather than model-dependent fits to specific scenarios,
we argue that an increase with energy of the positron fraction in cosmic rays, suggested by
several experiments at $E\agt 7\,$GeV, most likely requires a primary source of electron-positron pairs. We discuss the
possible alternatives, and find none of them plausible on astrophysical or particle physics
grounds. Further observational ways to test different scenarios are discussed.
\end{abstract}
\pacs{98.70.Sa}

\maketitle
\section{Introduction}
\begin{figure}[!htb]
\begin{center}
\begin{tabular}{c}
\epsfig{figure=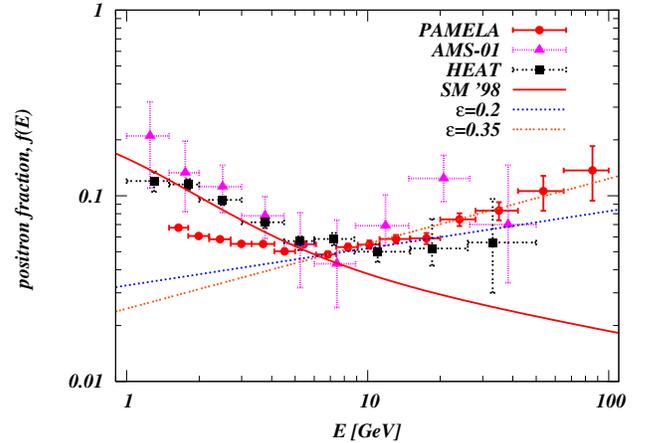,width=1.0\columnwidth}
\end{tabular}
\end{center}
\vspace{-0.9pc} \caption{The figure reports the positron fraction vs. energy measured by PAMELA~\cite{Adriani:2008zr},
 HEAT combined data~\cite{heat} and AMS-01~\cite{ams01}. A ``typical'' prediction based on purely secondary $e^{+}$ often used as a benchmark in the literature~\cite{Moskalenko:1997gh} and power-law curves $E^{\varepsilon}$ passing through PAMELA datum at 6.83 GeV with index $\varepsilon=0.2$ and $\varepsilon=0.35$ are also reported for illustrative purposes.}\label{plot}
\end{figure}

Since longtime, the study of the positron/electron ratio in cosmic rays has been recognized
as an important tool to constrain the production and propagation of energetic particles in the Galaxy and in the Solar System.
The PAMELA satellite detector, which began its three-year mission in June of 2006, is designed to measure (among other components) the spectra of cosmic ray positrons up to 270 GeV and electrons up to 2 TeV, each with unprecedented
precision~\cite{pamela}. Recently, the PAMELA
collaboration has presented first results of the measurement of
the positron fraction in the cosmic ray spectrum, which appears to begin climbing quite rapidly between $\sim 7$ GeV and 
100 GeV~\cite{Adriani:2008zr}. A similar trend was in fact also indicated 
by earlier experiments, including HEAT~\cite{heat} and 
AMS-01~\cite{ams01}, although with lesser statistical
significance and over a smaller dynamical range.  The behavior that seems to emerge in the positron fraction  is very different from that predicted for secondary positrons produced in the
collisions of cosmic ray nuclides with the inter-stellar
medium (ISM). The  situation is summarized in Fig~\ref{plot}. 
While unaccounted systematics in the measurements are in principle possible,  we think it is worth reviewing
what kind of physics may lead to such an energy spectrum; we shall argue that by far the
simplest and most likely (astro)physical interpretation is that an additional, primary source of high energy  positrons  exists. 

The positron fraction is defined as
\begin{equation}
f(E)\equiv\frac{\Phi_{e^{+}}}{\Phi_{e^{+}}+\Phi_{e^{-}}}=\frac{1}{1+(\Phi_{e^{-}}/\Phi_{e^{+}})}\,,
\end{equation}
where the fluxes $\Phi_i$ refer to the ones at the top of the atmosphere. Here and in the
following, we shall keep implicit the dependence of fluxes on energy $E$.
We re-cast the fluxes in terms of physically motivated
contributions, without loss of generality:
\begin{eqnarray}
\Phi_{e^{+}}&=&\cP+\cS\,,\\
\Phi_{e^{-}}&=&\cA+\cP+\cS-\cD=\Phi_{e^{+}}+\cA-\cD\,.
\end{eqnarray}
The term $\cA$ represents the component of primary electrons accelerated in addition to any  $e^{+}-e^{-}$ pairs;  this term includes (but is not necessarily limited to) primary electrons accelerated in a typical ISM environment where no pairs are present. The term $\cS$  represents
the secondary component of $e^{+}$ produced in hadronic cosmic ray collisions in the ISM. Note
that an analogous term exists for $e^{-}$ as well, which we denote $\cS-\cD$: there is indeed
a small deficit $\cD$ of secondary $e^{-}$ compared to secondary $e^{+}$ due to charge asymmetry in the cosmic ray and ISM nuclei population, which is proton-dominated.  Finally, we allow for a putative primary flux  $\cP$ of $e^{-}$-$e^{+}$ pairs, which is put to zero in typical predictions of $f(E)$. A $\cP$-term might be due for example to unaccounted astrophysical accelerators as pulsars or to a more exotic source  as dark matter (DM) annihilation. 
In terms of these components, 
\begin{equation}
f(E)=\left(2+\frac{\cA-\cD}{\cP+\cS}\right)^{-1}\,.
\end{equation}
First, one trivial observation: since from particle physics we know that $\cD\geq 0$, from the empirical datum $f(E)<1/2$ it follows that $\cA\geq \cD \neq 0$. This is not surprising, since it is well known  that accelerators of $e^{-}$ (i.e. electrons only, not pairs) exist in nature. Shocks on the background ISM at supernova remnants (SNRs, the most likely accelerators of Galactic Cosmic Rays) are the prime candidate in that sense; for modern predictions of the electron spectrum and an overview of past publications, see e.g.~\cite{Kobayashi:2003kp} and refs. therein. The low-energy behavior of $f$, typically found to decline from $\sim 0.1$ to $\sim 0.05$ up to $\sim 5\,$ GeV, has been measured since longtime. This range is influenced by time and charge-dependent solar modulation, which is likely responsible
for the differences among experiments. We shall neglect in the following solar effects which are irrelevant in the high
energy range we focus on here.
On the other hand, growing evidence has been collected in the recent years that  $f(E)$ might be rising at high energies, 
with the latest PAMELA data strongly favoring this observation. While one should wait for higher statistics and possibly an independent confirmation (in particular by AMS-02~\cite{AMS-02}), it is useful to classify the possible (astro)physical mechanisms leading to such an effect, a task we discuss
in Sec.~II. We will conclude that the only one appearing viable requires $\cP\neq 0$, which would imply the discovery of a new class of cosmic-ray sources (Sec.~III). 

\section{The necessity of a primary spectrum of $e^{+}$-$e^{-}$ pairs}
\subsection{Basic Arguments}
To prove the statement in the title of this Section, we shall adopt a ``reductio ad absurdum'' approach. Let us note that $f^\prime\equiv {\rm d} f/{\rm d} E>0$ implies 
\begin{equation}
\left(\frac{\cA-\cD}{\cP+\cS}\right)^\prime<0\,.\label{cond} 
\end{equation}
Let us now consider the {\it Ansatz}  $\cP=0$ and neglect $\cD$ for the moment (we shall see that this is justified, actually even a conservative assumption). Then, we should require  $(\cA/\cS)^\prime<0$ to produce a rise. We see now that this requires highly implausible astrophysical conditions.

In a very simple (position-independent) leaky-box model, the master equation for leptons simplifies into~\cite{Ginzburg:1990sk}
\begin{equation}\label{conteq}
\frac{\partial \Phi}{\partial t}= 
q(E)-\frac{\Phi}{\tau(E)}-\frac{\partial}{\partial E} \left[b(E)\Phi \right]\,,
\end{equation}
where $\Phi$ is the lepton flux, $q(E)$ the
initial/injection spectrum, $\tau(E)\propto E^{-\delta_e}$ an effective containment time, and
$b(E)\equiv-dE/dt\simeq \kappa E^2$ the energy-loss rate function, which at the energies
of interest (E$\agt 7\,$GeV) is dominated by synchrotron radiation and inverse Compton scattering. Let us assume spectra at the injection of the typical power-law form $\propto E^{-\gamma_e},\,E^{-\gamma_p}$ respectively for electrons and for protons (which, being the dominant hadronic component of cosmic rays and ISM, are the main responsible for secondary leptons). 
Protons suffer virtually no energy losses and obey an equation similar to Eq.~(\ref{conteq}) with $b(E)\to 0$, which yields for the steady state solution (${\partial \Phi}/{\partial t}=0$) a spectrum $\propto E^{-\gamma_p-\delta}$ with, a priori, $\delta\neq \delta_e$. The index $\delta$ is constrained from the nuclide ratio B/C to lie in the range $0.3\div 0.6$ (a review of recent cosmic ray experiments and their interpretation is provided in~\cite{Blasi:2008ch}). The convolution of this spectrum with the relevant cross-section is the source term for positrons, $q_{+}\propto E^{-\gamma_p-\delta}$ assuming an energy-independent inelastic cross-section and inelasticity. For the electrons,  one has simply $q_{-}\propto E^{-\gamma_e}$, plus a subdominant secondary contribution of
similar magnitude and shape of the positron one. The resulting spectra of primary electrons and secondary positrons at the Earth are thus respectively $\propto E^{-\alpha_{-}},\,E^{-\alpha_{+}}$ where  $\alpha_{-}=\gamma_e+\ell$,  $\alpha_{+}=\gamma_p+\ell+\delta$. Here $\ell$ symbolically represents the steepening due to diffusion and energy losses of leptons. For example, when one can neglect energy losses, $\Phi \propto q(E)\,\tau(E)$ and $\ell=\delta_e$; at sufficiently high-energy  (TeV range)  where energy-losses dominate it is easy to see that $\Phi \propto b(E)^{-1}\int dE^\prime q(E^\prime)$ and $\ell=1$. Independently of the value of $\ell$, in this simple model we end-up with $(\cA/\cS)\propto E^{-\gamma_e+\gamma_p+\delta}$ and thus Eq.~(\ref{cond}) requires $\gamma_p+\delta-\gamma_e<0$. This condition seems extremely hard to achieve, requiring wildly different  (by $\agt 0.6$)  source spectral indexes for protons and electrons. This would contradict the standard theoretical interpretation of the spectral difference between $p$ and $e^{-}$ observed at the Earth as due to different energy-loss properties, rather than intrinsic ones. Also note that the condition $\gamma_p+\delta-\gamma_e<0$ could not hold down to low energies, since the 
the sign of $f^\prime$ is negative around GeV energies. So, one should also invoke some spectral break in the injection electron spectrum placed {\it ad hoc} in the $\sim 7\,$GeV range. In summary,  insisting in the prior $\cP=0$ and requiring thus $(\cA/\cS)^\prime<0$ seems to imply: i) that our scenarios for the origin of Galactic electrons are wrong, requiring in turn either new sources or new acceleration mechanisms different from the proton ones; ii) some degree of fine tuning, in the sense that the energy at which $f^\prime$ changes sign would correspond to some spectral break in the electron spectrum.
These conditions appear way more extreme than allowing for primary sources of pairs, for which candidates (both astrophysical and exotic) do exist in the literature. 

On the top of the above considerations, there are empirical arguments which appear to disfavor this hypothesis. Defining $\Phi_{tot}\equiv\Phi_{e^{+}}+\Phi_{e^{-}}$, from Eq.~(1) one has
\begin{equation}
\frac{\Phi_{tot}(E_1)}{\Phi_{tot}(E_2)}=\frac{f(E_2)}{f(E_1)}\frac{\Phi_{e^{+}}(E_1)}{\Phi_{e^{+}}(E_2)}\,.
\end{equation}
Assuming {\it only} secondaries, the {\it hardest} spectrum theoretically possible for positrons derived in~\cite{Delahaye:2008ua} goes as $\sim E^{-3.33}$ above 10 GeV. In the same paper it is reported that the {\it softest} possible spectrum for $\Phi_{tot}$  fitting  (poorly, at 3 $\sigma$) the data compiled in~\cite{Casadei:2004sb} goes as $\sim E^{-3.54}$ in the same range. As a result, the maximal growth possible for the positron ratio---assuming only secondary positrons and no priors for the electrons---is
\begin{equation}
\frac{f(E_2)}{f(E_1)}\alt\left(\frac{E_2}{E_1}\right)^{0.2}\,.
\end{equation}
As illustrated in Fig.~\ref{plot}, this is insufficient to fully explain the rise suggested by the PAMELA data. More in general, this argument proves that a better determination of the electron energy spectrum might reveal an inconsistency of the $f(E)$ and $\Phi_{tot}$ data with a purely secondary origin of $e^{+}$, which does not resort to any theoretical considerations on the $\Phi_{e^{-}}$ flux.

\subsection{Possible loopholes: further discussion}
Although illustrated in a simple leaky box scenario, the conclusion that $\cP\neq 0$ is required appears robust. A subtle point in the considerations following Eq.~(5) is that we really need to compare the spectrum of the primary electrons with the one of cosmic ray nuclei at much higher energies (a factor $\agt 20$) since secondary leptons only carry a limited amount of the parent nucleus energy. Although a concavity of the spectrum would be naturally accommodated in non-linear acceleration
models~\cite{MD01}, there is no evidence that the (well measured) proton spectrum presents a noticeable change of slope around TeV energy, certainly not at the level of $\Delta \gamma_p\simeq 0.5$. 

Another way around the previous conclusion may be to consider a progressively rising role of Helium nuclei as source of secondaries. Still,  at energies around the TeV one should require a flux of Helium nuclei comparable to proton one and, at the same time, its spectral index harder than the proton one by an amount larger than $\delta$. Actually, some indications of a hardening of Helium spectrum has been claimed by the ATIC-2 collaboration~\cite{Panov:2006kf}. But its amount and the energy range where it happens appear insufficient to explain the beaviour of $f(E)$. For example, between 200 GeV and 1 TeV the flux ratio $\Phi_p/\Phi_{\rm He}$ varies by only $\sim 15\%$. To exclude this possibility, however, it would be important to compare the positron fraction and $p$ and He spectra measured with the same instrument. Preliminary results by PAMELA,
for example, do not support such an explanation since they show that both $p$ and He fluxes are well fitted with the same spectral index $\simeq 2.73$ up to $\sim\,$500 GeV~\cite{Boezio:FNAL}.

In~\cite{Moskalenko:1997gh}, the possibility was discussed that the average interstellar proton spectrum may be harder than the one measured ``locally''   by an index $\sim 0.15$, invoking both a better agreement with the diffuse gamma-ray spectrum and the HEAT data on the positron fraction. Note that even this ad hoc adjustment would be insufficient to explain the present evidence supporting $f^\prime>0$. Still, assuming that this argument is correct, one would expect two qualitative predictions: since the ``collecting cosmic ray volume'' in the Galaxy depends both on primary type and energy (via diffusion and spallation effects), a spatial non-universality of cosmic ray acceleration should reflect into a change of slope of a single species vs. energy, and of different species from one another. The TRACER collaboration has instead reported that for nuclei between Oxygen and Iron a single power-law index$\sim 2.7$ can fit all the data in the GeV to TeV energy per nucleon, with possible variations within $\sim 0.05$~\cite{TRACER}. We take here the agnostic point of view  that  the gamma-ray  ``excess'' is not understood at the moment (it might even be due to a calibration problem, see~\cite{Stecker:2007xp}), but note that it may be considered as well as an indication that additional sources exist contributing above the GeV range, which is {\it consistent} with the hypothesis of  additional primary emitters of cosmic ray positrons. 

One may further wonder if a rising positron ratio might be due to an unexpected  energy-dependent beaviour of the diffusion index;  from previous considerations and in the simplest case of  $\gamma_p=\gamma_e$, it would follow indeed $(\cA/\cS)\propto E^\delta$. For the sake of the argument and with a slight abuse, let us assume $\delta$ to be ``slightly'' energy dependent (this is not rigorous since the previous solution has been derived for a constant $\delta$.) The condition $(\cA/\cS)^\prime<0$ translates into the requirement that $\delta(E)$ declines with energy {\it faster} than $1/\ln(E)$ in the 0.1-1 TeV energy range of the parent nucleus producing the positrons. This is a relatively large effect, with $\delta$ dropping by at least a factor $\sim 3$ in a decade of energy to account for the rising $f(E)$. Even the proposal that  $\delta$ changes from $\sim 0.6$ above 10 GeV to $\sim 0.3$ at TeV energies (~\cite{Kobayashi:2003kp} and refs. therein) appears insufficient to account for the sign of $f^\prime$. On the other hand, this argument faces another difficulty: the featureless  power-law of CR protons would result from a fine-tuned compensation of the variation of $\delta(E)$ and the injection spectrum, which seems improbable, the two being unrelated. It is worth noting however that even this baroque scenario is testable empirically from high-energy B/C data.

Another approximation in the previous argument is the assumption of a constant cross-section. A rising  inelastic cross section would reflect in the secondary energy spectrum. Indeed, the inelastic cross section grows with energy (see e.g.~\cite{Kamae:2006bf}), but only logarithmically, by $\sim 1\%$ between 10 and 100 GeV and by $\sim10\%$ between 0.1 and 1 TeV, i.e. equivalent at most to a power-law of index $\sim 0.04$. This is more than one order of magnitude smaller than what needed to explain the positron feature.
 
Finally, let us come back to  $\cD$, or better  $\cD/\cS$. The above considerations are
{\it a fortiori} true if $(\cD/\cS)^\prime\leq 0$, i.e. if the relative difference between positrons and electrons remains
constant or declines with energy. Note that this function is mostly dependent on particle physics, apart for the convolution with the primary cosmic ray spectrum. If we take for example the
$e^{+}$ and $e^{-}$ yields for a proton power-law spectrum reported in  Fig.~12 of Ref.~\cite{Kamae:2006bf} one can conclude that: i) the secondary spectra are  slightly harder than the primary one, consistently
with the energy-dependence effect discussed above; ii)  the electron spectrum is
slightly harder than the positron one, i.e. $\cD/\cS$ is slightly decreasing with energy. Although a detailed study would be required to assess more quantitatively the uncertainties in this argument, it is clear that a sufficiently strong dependence of $\cD/\cS$ which might account for a rise in $f(E)$ appears out of question.\\{}\\

\section{Conclusions}
In summary, motivated by observational evidence and in particular by PAMELA data~\cite{Adriani:2008zr}, we have discussed under which conditions a rise in the positron fraction $f(E)$ can take place. Barring the case of systematics in the measurements,  we have analyzed the following hypotheses:
\begin{itemize}
\item[1)] ``Anomalous'' primary electron source spectrum.
\item[2)]  Spectral feature in the proton flux responsible for the secondaries.
\item[3)]  Role of Helium nuclei in secondary production.
\item[4)]  Difference between local and ISM spectrum of protons.
\item[5)]  ``Anomalous'' energy-dependent behaviour of the diffusion coefficient. 
\item[6)]  Rising cross section at high energy.
\item[7)]  High energy behavior of the $e^{+}/e^{-}$ ratio of secondaries in $p\,p$
collisions.
\end{itemize}
All of the above options seem to be at least strongly disfavored if not  already ruled out; nonetheless,  we have summarized the signatures associated with different explanations, and the way to test them observationally. Among the options listed which assume $\cP=0$, the one coming closer to a (very bad) fit to the data is nr. 1), which is not only disfavored by the data, but requires an ad hoc adjustment,  lacking at the moment an astrophysical model producing it.  We concluded that the most likely cause of the energy trend of the positron fraction is the presence of a primary flux of $e^{+}$, which---both in astrophysical and exotic models---are probably injected in the form of pairs (see~\cite{Coutu:1999ws} for an early review of possible primary sources).
Accepting this solution, one has at high energies $f(E)\simeq(2+\frac{\cA}{\cP})^{-1}$;
 from a rise at $E\agt 7\,$GeV one can further deduce that the spectrum of pairs is harder than the one of primary electrons, which is also a typical prediction in pulsar or DM annihilation/decay models.
In the opinion of the author, the positron spectral shape and normalization suggest pulsars as the most plausible responsible for the emission, a possibility which has drawn some attention lately~\cite{Hooper:2008kg,Yuksel:2008rf,Profumo:2008ms}. At very least, these objects should be seriously considered among the main actors of the high energy Galactic sky; perhaps they are also responsible for most unidentified Galactic gamma ray sources,  as recently supported by the discovery by FERMI of a pulsating gamma-ray emission from the SNR CTA 1~\cite{Kanbach:2008nz}.\\
{}\\
{\bf Acknowledgments} 
The author would like to thank P. Blasi for the initial suggestion to explore the
topic discussed here, and P. Blasi and F. Donato for reading the manuscprit and useful comments.


\end{document}